\documentclass{article}
\usepackage[english]{babel}
\usepackage[letterpaper,top=2cm,bottom=2cm,left=3cm,right=3cm,marginparwidth=1.75cm]{geometry}
\usepackage{amsmath}
\usepackage{amssymb}
\usepackage{graphicx}
\usepackage{stmaryrd}
\usepackage[colorlinks=true, allcolors=blue]{hyperref}
\usepackage{graphicx,stmaryrd,yfonts}
\title{Consciousness As Entropy Reduction (Short Version)}
\author{{\small Yifeng Chen, META Institute, School of CS, Peking University, Beijing, China ({\tt cyf@pku.edu.cn})}\\
{\small J.\,W.\,Sanders, DISEI, ECNU, Shanghai, China ({\tt sanders@sei.ecnu.edu.cn})}}
\date{}
\newtheorem{Def}{Definition}

\newtheorem{Thm}{Theorem}

\newtheorem{Prop}{Proposition}

\begin{document}
\maketitle
\newcommand{\Df}{\ \widehat{=}\ }
\newcommand{\hide}[1]{}
\newcommand{\msf}{\mathsf}
\newcommand{\mi}{\mathit}
\newcommand{\st}{\,\cdot\,}

\begin{abstract}
A model of consciousness is proposed which, having a logical basis, lends itself to simulation using a simple mathematical model called Consciousness as Entropy Reduction (CER). The approach has been inspired by previous models such as GWT, IIT and an earlier less mainstream model called ``Feature Map'' in Psychology. 
CER considers the contents of consciousness and subconsciousness as \textit{scenarios}: a vector of patterns (or features) on various ``channels'' (or feature locations). In CER, a feature map itself is not consciousness but only the input \textit{scenario} into a world of possible subconscious \textit{scenarios} from which the conscious \textit{scenario} (i.e., conscious experience) is chosen. Essentially, it creates an internal simulation of the outside world. Solving problems in simulation internally as a ``thought experiment'' is obviously more economical than doing experiments in a real environment and lends itself to adaptability and hence is a major evolutionary advantage. 
CER also has connections with the Hopfield model in artificial neural networks.
\hide{The approach is demonstrated on three examples: binocular rivalry, Libet's experiment, and the formation of higher-order consciousness.}
\end{abstract}

\section{Introduction}

Although familiar to all, the notion of consciousness lacks a consensual definition. Part of the difficulty is attempting a rational explanation of subjective experience. Another is the confusing variety of uses of the term, interpreted differently in different contexts. 

In this paper, we adopt the terms ``unconscious'', ``subconscious'' (or preconscious) and ``conscious'' used by Sigmund Freud. A knee-jerk reaction is unconscious. Consciousness does not trigger the responding actions but can still observe and analyse the fast reaction as a sequence of fast-occurring events outside of consciousness. Precision control for movement with posture feedback is also unconscious, although some previous models do not distinguish these terms and may treat such a system (e.g.,~predictive-coding systems~\cite{coding}) to be conscious. Subconsciousness typically includes competing desires. For example, an individual must balance its survival desire, sexual desire and offspring-protecting desire in its response to a current danger. It is advantageous to run multiple options of recognitions or actions and choose the most suitable as the conscious. In short, consciousness is the tip of an iceberg with also  subconsciousness and unconsciousness.
\hide{in various disciplines with contrasting approaches, from Neuroscience, Philosophy of the Mind, Computer Science to Mathematics. Furthermore nowadays the quest includes the meaning of consciousness for entities other than humans, including AIs. The concept seems to have been confused rather than clarified by its relationship to things like self awareness and intelligence. Instead many architectures have been proposed supposedly to account for consciousness.}

In appreciation of the context, this paper proposes a mathematical model, Consciousness as Entropy Reduction (CER). The assumptions of the model are related to previous models such as Global Workspace Theory (GWT~\cite{gwt}), Integrated-Information Theory (IIT~\cite{iit, iit2}) and an earlier less mainstream model called Feature Integration Theory~\cite{fit} from psychology. Like the latter, CER considers the contents of consciousness and subconsciousness in terms of scenarios (or maps).  A {scenario} consists of  a number of ``patterns'' (or features) on various ``channels'' (or feature locations). 

The subconscious world is like an ambiguous image from which one conscious scenario arises as a clarification of the image by utilising genetically inherited instincts as well as previously learned experiences. The ability to run multiple optional hypotheses has obvious evolutionary advantage in an ecological environment, for example, with widespread camouflage and mimicry. 
The CER mechanism of choosing a conscious scenario from subconscious scenarios is 
specified abstractly\hide{as rather than algorithmically [CYF: gradient descent can be viewed as algorithm]} as an interface called the S2C (i.e., subconsciousness-to-consciousness) interface. 

The presence of S2C in humans is supported by the experiment of binocular rivalry~\cite{bino}: when the tested is presented with a different picture to either eye at the same spot in the field of vision, the tested reports seeing both whole pictures in an alternating order.\hide{The more visible and distinctive are the pictures, the more frequent alternations occur.} That means the brain always chooses a whole pattern instead of a montage in each conscious experience. This experiment also shows that human consciousness is channel-sensitive.  Competition exists among subconscious patterns in the same channel (e.g., at the centre of the vision field).

S2C in CER has similarity to the model GWT in which the content of a global workspace is chosen from contents of local workspaces. The choosing mechanism is unspecified in GWT. One popular suggestion~\cite{blum} is that the subconscious local workspaces compete through some comparing structure, and the ultimate winner becomes the global content.\hide{One advantage of using a specification is that it leaves open many implementations and simulations, though constrained by the specified properties.}

The mechanism S2C in CER is more relatable to a Hopfield network~\cite{hopfield} in  artificial neural networks. A Hopfield network that starts in some initial vector state changes its state in the direction of reducing some energy function (usually defined as a Euclidean variance function). Eventually, a network with properly interconnected weights always reaches one of the energy-minimal stable states. In some sense, one (among many) stable global state  (or scenario) in memory arises to be close to the initial vector state by gradient descent of the energy function. 

Subconscious scenarios in CER are represented as a probabilistic distribution of patterns over channels. Unlike Hopfield networks, the energy function in CER is a kind of entropy function that reflects instinctive preferences on how channels are correlated in integration. The S2C interface is assumed to be effective and efficient in reducing  entropy in a  gradient-descent process until a zero-entropic distribution is ``determined'' as the chosen conscious scenario. The Markovian nature of the gradient-descent procedure bears some similarity to the IIT model. But unlike assumptions in IIT, CER depends on the mechanism that chooses the conscious scenario out of numerous subconscious scenarios. This is more in line with the GWT model.   Species with the S2C mechanism are more flexible in decision-making and hence adapt to a fast-changing environment better. 

The reduction of entropy exists in various kinds of information processing. For example, when seeing a partial imagine that may have different interpretations, the brain automatically tries out possible supplement details and determines the likely interpretation in an energy-consuming process. IIT is defined to reflect the reduction of entropy caused by the correlation (or integration) among different parts of a system. As IIT is defined on the dynamics of a system, the positive entropy of the overall system cannot be reduced to 0. Although GWT has not been described as entropy reduction, but it actually corresponds to a typical kind of entropy reduction: from positive-entropy distribution of local workspace contents to 0-entropy global content. Despite the rivalry between the two theories, entropy reduction is their fundamental common ground. The main difference is that the former does not reduce entropy to zero, while the latter does.

The S2C choice mechanism alone does not explain all we know about consciousness, which is also related to the notion of ``qualia'' or subjective experiences. Two colours red and blue may be perceived  symmetrically but they are ``felt'' differently. This is known as the ``hard problem'' of consciousness~\cite{hard}. In fact the content squeezed out of subconsciousness may not be fully noticed consciously at the time. GWT proposes a spotlight mechanism to describe the highlights of the noticed content in the global workspace as a stage. The items in the spotlight become the subjective experience, but GWT does not specify how the spotlight chooses the items on stage. Which items to receive the spotlight is an internal choice. A similar choice mechanism is also required for recurring dreams with a variable storyline of forking plots. Such a dream consists of sequentially ordered conscious steps in a chain. Each step of conscious experience is a chosen scenario from numerous subconscious scenarios. 

Dreams are interesting as the brain works internally with almost no physical interaction with the environment. The dream of a primitive species may contain only flashy snapshots, while an advanced species that can dream a storyline will likely also be capable of planning and logical reasoning as some kind of ``slow thinking''~\cite{kahneman,bengio}.  We humans  have a vivid ``feel'' about the scenarios in a dream, although when we wake up and try to recall it, the dream seems to be much less detailed than a real experience. That suggests our feel is formed upon internally generated patterns, not directly from the real sensory inputs, whether awake or dreaming.

The ability to sustain an internal chain of experiences varies drastically among different species. Albatross is seem to ignore their struggling chick just blown out of the nest by storm and refuse to feed it. On the other hand, experiment~\cite{apes} shows that gorillas can figure out the steps from the problem to its solution entirely in their mind before acting on it. Humans are perhaps the best at sustaining a long internal chain of thought experiments for planning and story-making, for example, with hours in problem solving or even days in meditation. Solving problems internally as thought experiments is obviously more economical than conducting real experiments in a physical environment. 

The primitive CER model is extended with an information pathway called C2S (i.e., consciousness-to-subconsciousness feedback) that feeds conscious patterns back to the subconscious world so that imagined or dreamed contents can join subconsciousness for further S2C processing without interacting with the physical environment first.
S2C and C2S in combination support not only
richer imagination, more sophisticated dreams, and better planning/problem-solving but also ``high-order'' thoughts~\cite{HOT}, which are typical characteristics for more evolved forms of consciousness. 

Humans can process not only concrete patterns of real objects ifrom the physical environment but also subjective concepts such as self, self awareness/consciousness and subjective experiences, which become internally generated abstract patterns in CER. For example, after attending similar meetings repeatedly, seeing the video slip of a similar meeting scene from the perspective of a usual seat infers the attendance of self, which may next infer the subjective feel about self's presence. Humans are self-conscious and have various internal patterns related to the sense of self. The mechanism of self consciousness must, of course, rely on both S2C and C2S to work.

In summary, the proposed CER model makes the following the assumptions:
\begin{enumerate}
    \item Consciousness is channel-dependent. Multiple patterns compete in a channel, and at most one pattern appears in  that channel in each conscious experience. Consciousness is a sequence of conscious experiences. 
    \item The integration of different channels according to past training and inherited instincts is described in the form of entropy that is reduced in the fastest way until  a conscious experience emerges from subconscious scenarios. All internal choices go through such integration and entropy reduction.
    \item Attention, planning, high-order thoughts and storyline dreaming all require a feedback mechanism to project conscious patterns back into the subconscious world. 
    \item Self consciousness is explained as internal imaginative patterns in reflection through feedback. 
\end{enumerate}

In CER, we propose that species with S2C are {\bf primitively conscious} with conscious scenarios, primitively conscious species with C2S are {\bf high-order conscious} with storyline experiences, and high-order-conscious species aware of self are {\bf subjectively conscious} with subjective experiences. 

\hide{
+ some comments about the math tools of the model needed
}

\hide{
Despite the simple setup of the CER model, it naturally gives rise to some known phenomena in psychology. 

+ mention binocular rivalry again

+ the sea of subconscious world supports a kind of short-term memory as patterns linger in it for some time. this helps establishing relation between patterns occurring in a short lapse of time

+ The Libet experiment shows that neural activities of action tend to occur hundred milliseconds ahead of the person's report on ``deciding to act''.  This is simulated in CER as a delayed conscious report that is formed by analysing previously-occurring events and actions (as patterns). Such interpretation is unrelated to the debate about free will that often surrounds Libet's experiment. It only points to the fact that the experience of ``making a decision'' is not necessarily the actual decision-making. 
}

\section{CER Model}
\subsection{Single-Channel Distribution of Patterns}
Consider a probabilistic distribution $\mathit{vision}=\{{\mathtt{house}}\mapsto 0.5,\, {\mathtt{face}}\mapsto 0.5\}$ for a single channel of vision. Either pattern has half the likelihood. 
All other patterns have probability 0. If a pattern does not occur in a channel, it is simply represented as zero probability in the distribution. For example, an imaginative pattern for a certain feeling exists only internally and thus always has zero probability in a vision channel. Other patterns such as sounds in hearing and subjective feelings never occur in a vision channel and hence also have zero probability.

In general each single-channel distribution $x$ is a function $x{:}P\rightarrow [0,1]$ with a sum to 1 where $P$ is the set of all patterns. A special pattern {\tt void} denotes the absence of any particular pattern in the channel. A distribution is {\em determined} (e.g., $\{{\mathtt{void}}\mapsto 1\}$), if it is a point-mass distribution with exactly one pattern of probability 1 and all other patterns of probability 0. 

We also use $x_S=\sum_{r\in S}\,x(r)$ to denote the sum probability of a subset $S\subseteq P$ of patterns. We thus have $x_\emptyset=0$ and $x_P=1$. The probability $x(p)$ of a pattern $p$ is equal to $x_{\{p\}}$.

\subsection{Scenario of Multiple Channels }
Advanced systems of information processing will correlate information from different channels such as vision and hearing in an integrated scenario.
Let us introduce  another channel of internal feels and a distribution $\mathit{feel}=\{\mathtt{small}\mapsto 0.7,\, \mathtt{big}\mapsto 0.3\}$ with internal patterns {\tt small} and {\tt big} associated with the channel. Then the following is a joint distribution of the two channels {\it vision} and {\it feel}:
\[\mathit{vision}\times\mathit{feel}=\{(\mathtt{house},\mathtt{small})\mapsto 0.35,\,(\mathtt{house},\mathtt{big})\mapsto 0.15,\,(\mathtt{face},\mathtt{small})\mapsto 0.35,\,(\mathtt{face},\mathtt{big})\mapsto 0.15\}.\] 
In general, a scenario probabilistic distribution of $n$ channels is a function $x{:}P^n\rightarrow [0,1]$ with a sum of~1. The channels are numbered from 1 to $n$. 
A simple way to conjoin two distributions $x{:}P^n\rightarrow [0,1]$ and $y:P^m\rightarrow [0,1]$ is their Cartesian product: $(x\!\times\! y):P^{n+m}\rightarrow [0,1]$ in which $(x\!\times\! y)_{\{(r,s)\}}=x_{\{r\}}\,y_{\{s\}}$ for any $r\in P^n$ and $s\in P^m$. Cartesian product reflects independent conjoining of two distributions with maximum conditional entropy. 
A scenario distribution $x$ (e.g.,~$x=\mathit{vision}\times\mathit{feel}$) is ``fully unintegrated'' if it is the Cartesian product of $n$ single-channel distributions: $x=\prod_i^n\, x_i$. A determined distribution is also a special unintegrated distribution.
For example, the determined scenario distribution $\{(\mathtt{house},\mathtt{small})\mapsto 1\}$ corresponds to a scenario $(\mathtt{house},\mathtt{small})$ with the  pattern {\tt house} in the first channel and the pattern {\tt small} in the second. 

In the CER model, we assume that a subconscious distribution is fully unintegrated, while a conscious distribution is determined. Integration and correlation only occur when the nervous system tries to clarify the contents in the subconsciousness and chooses a suitable conscious interpretation out of the subconscious world. 

\subsection{Entropy}
Entropy is a powerful mathematical tool for probabilistic modelling. In this paper, we adopt a convenient entropy term called ``refusal entropy''.

\begin{Def}[Refusal Entropy]
     The refusal entropy of a probability $a$ is defined: $R(a)=a\log \frac{e}{a}$ where $e$ is the natural constant, and the base of logarithm is 2.
\end{Def}
The first and second derivatives of refusal entropy are $\log \frac{1}{a}$ and $-\frac{\log e}{a}\leqslant 0$, respectively, satisfying monotonicity, convexity and strictness $R(0)=0$ (as  a usual assumption in information theory). For a binary distribution of two probabilities $a$ and $b$ such that $a+b=1$, the usual definition of entropy  $H(a,b)=a\log \frac{1}{a}+b\log \frac{1}{b}$ is equal to $R(a)+R(b)-\log e$, and we also have $\frac{\partial H(a,b)}{\partial a}= R'(a)$ and $\frac{\partial H(a,b)}{\partial b}= R'(b)$. The refusal entropies to the  usual definition as they are monotonic and hence convenient for representing the rules for inter-channel integration.

Changing $a$ and $b$ by performing gradient descent in the opposite direction of the derivatives $(R'(a), R'(b))$ and, at the same time, keeping the sum unchanged will eventually reach 0 or 1, except for an unstable balance point with a zero first derivative.

The usual entropy in Shannon's information theory has a similar definition as the refusal entropy terms:

\begin{Def}[Shannon's Entropy]
     $\mathbf{H}(x)=\sum_{r\in P^n}\, R(x_{\{r\}})-\log e$ where $x{:}P^n\rightarrow [0,1]$ is a scenario distribution.
\end{Def}

\begin{Prop} If $x$ and $y$ are two scenario distributions, then $\mathbf{H}(x\times y) =\mathbf{H}(x)+ \mathbf{H}(y)$.
\end{Prop}
For example, we have $\mathbf{H}(\mathit{vision})=1$, $\mathbf{H}(\mathit{feel})=0.881$ and 
$\mathbf{H}(\mathit{vision}\times \mathit{feel})=1.881$ for unintegrated combination of the two channels.
In comparison, the following integrated scenario distribution with the same projected single-channel distributions has lower Shannon's entropy:
\[\mathbf{H}(\{(\mathtt{house},\mathtt{small})\mapsto 0.3,\,(\mathtt{house},\mathtt{big})\mapsto 0.2,\,(\mathtt{face},\mathtt{small})\mapsto 0.4,\,(\mathtt{face},\mathtt{big})\mapsto 0.1\})=1.846.\] The total entropy is lowered because of the correlation between channels.

\subsection{Weighted General Entropy}
We further generalise the notion of entropy to incorporate rules that can integrate multiple channels.  Consider a rule that ``seeing a {\tt house}  leads to feeling {\tt big}''. This rule permits ``seeing a {\tt face}'' and ``seeing a {\tt house} with feeling {\tt big}'' but prohibits ``seeing a {\tt house} without feeling {\tt big}''. 
This can be captured as refusal entropy $R(x_{\{(\mathtt{house},p)\mid p\neq \mathtt{big}\}})$, which is zero for the probabilities of the allowed scenarios and positive  for the prohibited scenarios. Modifying the distribution by reducing this entropy will reduce the probability of prohibited scenarios. 

In general, the entropy of a scenario distribution has the following form:

\begin{Def}[Weighted General Entropy]
     $\mathbf{E}_w(x)=\sum_{S\subseteq P^n}\,w_S\, R(x_S)$ where $w_S{\geqslant} 0$ is the weight of the entropy of the subset $S$ of scenarios.
\end{Def}
The higher is the weight $x_S$ of a subset $S$, the more pressure the entropy  puts on  the distribution to reduce that entropy and avoid the scenarios in $S$. The weight function $w$ from every subset $S$ to its weight value uniquely determines the strength of the entropy term $x_S$ that prohibits scenarios in $S$. The condition $w_S=0$ indicates the absence of such a rule.\hide{ Every refusal entropy term is convex, so is the weighted general entropy. A non-trivial weighted entropy term has a unique symmetric point with maximum entropy.}

The entropy term $\mathbf{E}_w(x)$ is the general form of many interesting entropy definitions. 
For example, Shannon's  entropy term is a special case of weighted general entropy.

\begin{Prop} Equation $\mathbf{E}_w(x)= \mathbf{H}(x)+\log e$ holds where $w_{\{r\}}=1$ for $r\in P^n$ and $w_S=0$ for other subsets $S$ of scenarios.
\end{Prop}

\begin{Def}[S2C Interface]
The S2C entropy term is defined:
$\textstyle \mathbf{SC}_{w,\epsilon}(x)\ =\ \mathbf{E}_w(x) \;+\; \epsilon(x)$
where the random noise term $\epsilon(x)$ is an infinitesimal zero-sum vector with infinitesimal non-zero values and non-zero derivatives. 
\end{Def}
Randomness is needed to break symmetry of $x$ when the derivatives are all 0 mathematically. 

\subsection{Entropic Gradient Descent}
Every conscious scenario is a 0-entropic determined scenario distribution that corresponds to a combined scenario ``squeezed'' from the subconsciousness. The number of all possible conscious scenarios is easily astronomical. For example, with $n$ channels and $m$ patterns, the number is $n^m$. Since the amount of entropy reduced from an initial distribution to every 0-entropic conscious scenario is the same, the question is which scenario should be the chosen conscious scenario? 

It is unknown how the brain exactly minimises entropy, but it is reasonable to assume that, by evolution, the brain becomes effective and efficient in entropy reduction. Mathematically, the fastest direction to reduce an entropy function is determined by the partial derivatives of the entropy function. 

We use $\nabla \mathbf{F}(x)$ to denote the vector of partial derivatives for some entropy term $\mathbf{F}(x)$:
$(\nabla \mathbf{F}(x))_r= \frac{\partial \mathbf{F}(x)}{\partial x_{\{r\}}}$. 
The distribution $x$ should be modified in the opposite direction of $\nabla \mathbf{F}(x)$ to reduce entropy. Note that the modifying vector must always be normalised to have a sum of 0 to keep the modified  distribution remaining a distribution.

This is illustrated in a simple case $\mathbf{F}(a,b)=a\log \frac{e}{a}+b\log \frac{e}{b}+\epsilon(a,b)$ of two states where $a+b=1$. We ignore $\epsilon$ for the moment. Then we have the derivative vector $\nabla \mathbf{F}(a,b)=\left(\frac{\partial \mathbf{F}(a,b)}{\partial a},\,\frac{\partial \mathbf{F}(a,b)}{\partial b}\right)=\left(\log \frac{1}{a},\,\log \frac{1}{b}\right)$. To maintain the sum of probabilities unchanged in the distribution, the derivative vector needs to be normalised to 0 sum (by subtracting its average) as  $(\frac{1}{2}\log \frac{b}{a},\,\frac{1}{2}\log \frac{a}{b})$. Gradient descent following the opposite direction of the normalised vector will end in either $a=0$ and $b=1$ if $a< 0.5$, or $a=1$ and $b=0$ if $a> 0.5$. 
The unstable balanced point $a=0.5$ corresponds to the unique maximal entropy with 0 derivative. An infinitesimal perturbing noise $\epsilon(a,b)\neq 0$ is needed to break the symmetry for the gradient descent to work in either direction. Every vector dimension of the distribution  corresponds to a scenario, and its probability must be kept non-negative.  

A general gradient-descent process with multiple scenarios follows the opposite direction of normalised derivative vector. Once a probability  dimension reaches 0, further modification to that dimension will be ignored. Eventually the process terminates in a determined distribution. The corresponding scenario  will be the new conscious scenario at that moment.

\begin{Thm}[S2C Gradient Descent]
    The gradient descent of a non-zero general entropy always terminates in one of the determined distributions.
\end{Thm}
The proof follows the convexity of the weighted general entropy and the assumption about $\epsilon$.

Note that gradient descent of CER has another key difference from a usual (symmetric) Hopfield network besides  the different energy functions. The number of dimensions in a Hopfield network normally corresponds to the size of the system, while that number in CER is the number of all possible scenarios, which is exponential to the number of channels.

\subsection{Simulating Gradient Descent}
The advantage of introducing a mathematical model is to perform tests and experiments with the model. The technical challenge of simulating gradient descent in discrete steps is to keep the distribution always valid after every modification. 

The sum of a distribution after modification must be kept 1. That means the modifying vector must be normalised to zero sum for those scenarios with non-zero probabilities. This is achieved with an operator $y\ast S$ to denote normalisation of a vector $y$ for scenarios in the subset $S$ satisfying: $(y\ast S)_{\{r\}}=y_{\{r\}}-\sigma$ for $r\in S$ and $(y\ast S)_{\{r\}}=0$ otherwise. The constant $\sigma$ is such that  $(y\ast S)_{S}=0$.

\begin{Def}
The modification operator $x\pm y$ of a distribution $x$ with vector $y$ is defined as the following equation:
\[x \pm  y\ =\ (x+a\, (y\ast S))\;\pm\;  (1{-}a)\, (y\ast S)\]
where the subset $S$ satisfies $x_{\{r\}}=0$ iff $r\in S$, and  the parameter $a$ is small enough to keep  $(x+a\, (y\ast S))_{\{r\}}\geqslant 0$ for every $r\in S$.
\end{Def}
The definition implies an iterative  method to calculate the modification and keep the modified vector remaining a probabilistic distribution with a sum of 1.

We define the computer simulation of gradient descent from an initial distribution scenario $x$:
\[\begin{array}{ll}
    x^{(0)}\;=& x\pm  \epsilon \\
    x^{(1)}\;=& x^{(0)} \pm  \rho\, \nabla \mathbf{SC}_{w,\epsilon}(x)^{(0)}\\
    & \cdots\cdots \\
    x^{(n+1)}\;=& x^{(n)} \pm  \rho\, \nabla \mathbf{SC}_{w,\epsilon}(x)^{(n)}\\
\end{array}\]
where $\rho<0$ indicates the direction and the speed of gradient descent for entropy  reduction. We will use $\mathit{SC}_{w,\epsilon,\rho}(x)$ to denote the limit distribution's corresponding (conscious) scenario.
The design of $\pm$ is such that the set of probability-0 scenarios keeps expanding until only one scenario's probability is 1 in a determined distribution. The scenario is the conscious scenario $\mathit{SC}_{w,\epsilon,\rho}(x)$. 

\subsection{Primitive CER}
In CER, the subconscious distribution keeps tracking the external changes. The conscious scenario is retrieved by performing gradient descent on some weighted general entropy.

Let $v[t]{:}P^n\rightarrow [0,1]$ denote external\footnote{Unconscious reactions as well as sensory posture are also deemed external to the S2C interface. } unintegrated distribution of scenarios at time $t$. For those channels without any pattern, the special pattern {\tt void} will have a high probability. 
Let $x[t]{:}P^n\rightarrow [0,1]$ denote the unintegrated subconscious distribution, which will track the external distribution. The tracking mechanism is mathematically represented as a linear combination \[x[t]=\lambda\, v[t-1] + (1-\lambda)x[t-1]\] between the distributions where $\lambda$ is a factor for the linear combination. This allows external scenarios to be ``accumulated'' in the subconsciousness as plausible options. 
The next conscious scenario corresponds to the limit distribution of gradient descent: $\mathit{SC}_{w,\epsilon,\rho}(x[t])$. 

Species with such a primitive form of consciousness would be able to utilise  inherited instinctive preferences and past experiences to decide whether to treat an object as food or as danger instead. The retrieved conscious scenario not only clarifies the  recognition of external patterns but also triggers actions in response to the input. The species is only conscious of its own actions by observing the actual effect of the actions in the physical environment.

Primitive consciousness helps determine individual actions in fast-thinking reactions. For example, to form a plan with  steps S1, S2 and S3, a primitively conscious individual can take advantage of the past experiences, choose a possible action for the first step S1 and watch its effect to the environment. If S1 and S2 are repeatedly tested in the order, a joint experience of S1S2 may be established so that the individual learns the effect of S1 followed by S2 and can respond to an external stimulus by performing both S1 and S2 in order. A species with primitive consciousness is still able to learn a procedure (e.g., S1S2S3) by repeating S1S2 and S2S3 many times. Humans sometimes use this bottom-up technique to train in sports or martial arts.

\subsection{Higher-Order CER}
Experiments of Gestalt psychology, on the other hand, show that advanced species like Gorillas can think a while and perform all S1, S2 and S3 as a whole new plan without accumulating real experiences of either S1S2 or S2S3.  Real experiments are substituted by ``thought experiments'' with multiple steps.
Internalising a chain of S2C steps is not supported by the primitive CER model.

To support an S2C chain, a neural pathway is needed to feed the output conscious scenario back into the subconscious world so that the current conscious scenario can be subsequently analysed  in the following S2C steps. 

In CER, we assume, the feedback will change the subconscious probability of the conscious pattern in the same channel so that {\tt void} will have 0 probability if the current pattern is a fresh one in the channel, otherwise the probability of the current pattern will decay slightly. The following definition captures this assumption.

\begin{Def} [Channel Feedback]
    If the previous pattern is $p$ and the current pattern is $q$, then the feedback update $\mathit{cs}_\beta(x, p,q)$ to a single-channel distribution $x$ is defined so that $\mathit{cs}_\beta(x,p, q)_{\{\mathtt{void},q\}}=x_{\{\mathtt{void},q\}}$,     $\mathit{cs}_\beta(x,p, q)_{\{\mathtt{void}\}}=0$ if $q\neq p$, or 
    $\mathit{cs}_\beta(x,p, q)_{\{q\}}=\beta\, x_{\{q\}}$
    if $q= p$ where $\beta<1$ is a decaying factor.
\end{Def}

\begin{Def}[C2S Feedback] If a scenario distribution $x$ is an unintegrated product $x=\prod_i^n x_i$ of some single-channel distributions $x_i$, then the feedback update  $\mathit{CS}_\beta(x, r,s)=\prod_i^n \mathit{cs}_\beta(x_i,r_i,s_i)$ where $r,s\in P^n$ are the previous and current conscious scenarios, respective.
\end{Def}

The subconscious world denoted as $x$ not only follows the external input distribution $v$ but also takes in the influence of the conscious scenario as C2S feedback. The extended procedure is defined as follows:
\[\begin{array}{rl}
     y[t]=& \lambda\, v[t-1] + (1-\lambda)x[t-1] \\
     r[t] =& \mathit{SC}_{w,\epsilon,\rho}(x[t])\\
     x[t] = &\mathit{CS}_\beta(y[t],\,r[t-1],\,r[t])\\
\end{array}\]
The species with such a feedback mechanism will be able to test the effect of conscious actions in their mind and form a plan with consecutive steps before actually performing the first suitable action. This ability will also give rise to abstract patterns such as concepts and self for high-order consciousness.

\section{Discussions and Conclusion}
A species that can dream (or imagine) will suffice to have consciousness. There is evidence that at least some reptiles have dream-related sleep patterns~\cite{dream}. This puts the first appearance of consciousness in evolution somewhere as a common ancestry to birds and mammals. Primitive forms of consciousness may have existed long before that, while sophisticated  behaviours such as the instinct to care for young offspring after birth are evolved afterwards. For example, a mammal must balance its survival desire, sexual desire and offspring-protecting desire in its response to a current danger. The balancing ability obviously can benefit from the S2C mechanism. 

Humans often think of consciousness as one concept, but we actually refer to different forms of consciousness in different contexts. This is why confusion and disputes set in before we can reach any common definition. In this paper, we propose three levels of consciousness definitions. Before we decide whether a specific species is conscious, the referred  level of consciousness must be clarified.

It is perceivable to speculate that the proposed three levels of consciousness have evolved into existence at different stages of evolution: primitive consciousness, perhaps, in an environment with widespread camouflage and mimicry, high-order consciousness in a complex environment where repeating stepwise conscious actions becomes too expensive and subjective consciousness when recognising roles, identifying self and empathising with others become important in a social setting. 

The evolution of consciousness has been in the direction of increasing flexibility and internalisation. Adding more flexibility to learning is an effective evolutionary strategy to enhance lifespan epigenetic adaptivity of a species, which takes the pressure off the needed genetic adaptation in a fast changing habitat. Performing planning and thought reflection inside the brain achieves a similar advantage to what computer simulation does in scientific research. It is simply much more cost-effective to simulate than doing the real physical experiments. Thus more advanced species tend to internalise more aspects of information processing in the nervous system. 

Entropy reduction, as a mechanism, is expensive to integrate (for channel correlation), to run (with significant energy consumption), to develop (before adulthood)  and to evolve (as an evolutionary invention). It is therefore reasonable to assume that there is just one mechanism for S2C, which is used and re-used in subsequent evolutionary steps. High-order thoughts are the result of causally related internal thoughts being folded around the  S2C mechanism in iterative steps, just like the role of loops in programming languages. This is a crucial difference of CER from other models in which the structure of high-order thoughts is often believed to reflect the actual neural structure.

The set of patterns in any species is limited for sure. For example, we do not have experiences of wifi's electromagnetic signals. Most animals may not have patterns related to self. This explains why we do not know ``what is it like to be a bat''. It is unclear whether new patterns can be created in lifespan, as the mechanism of long-term memory is unclear. The basic model of CER needs to be extended or even modified to incorporate more mechanisms.

In some way, CER is like a mathematical game on top of which computer-simulation games can be set up and tested against psychological experiments. 
This means the model is falsifiable if all plausible setups of simulation fail to capture the findings of a psychological experiment.
Even if the simulation accurately corresponds to the psychological experiment, it only means the simulation may have captured the inner working of the mind. 
It does not mean the simulation itself is conscious. Ultimately whether a system is ``conscious'' depends on the adopted definition of consciousness. If the definition insists on the system being biological, then any non-biological system is  not conscious by definition.


\hide{New para: In fact the content squeezed out of subconsciousness may not be fully noticed consciously at the time. For example eidetic memory is the ability to recall fairly precisely a pattern (visual or aural), at least for a brief period of time, after seeing it only once [?], whilst hyperthymesia is the ability to recall an abnormally large number of experiences in vivid detail [?]. Neither is common. The former is demonstrated by 2 to 10 percent of children in the US between six and twelve but fades in adulthood. Of the latter there are currently fewer than 100 cases documented internationally. Both eidetic memory and hyperthymesia indicate that the human brain is capable of storing patterns for a short time enabling them to provide later conscious phenomena not originally perceived.

+ ON second thoughts, that para I sent should not refer to hyperthymesia. Here's an update: In fact the content squeezed out of subconsciousness may not be fully noticed consciously by the observer at the time. For example eidetic memory is the ability to recall fairly precisely a pattern (visual or aural), at least for a brief period of time, after seeing it only once [?]. It is not common, being possessed by about 2 to 10 percent of children in the US between the ages of six and twelve, and fades in adulthood. It allows the child to make observations about the pattern it had not made initially, so indicates that the human brain is capable of storing patterns for a short time enabling them to provide later conscious phenomena not originally perceived.

+ NDTM equivalent. as we can reduce SAT to CER (with each proposition term encoded as an entropy term).

+ void probably has small weight and easily reducible

+ not attention but notice. conscious content may not be noticed in the next round of second-order consciousness. such patterns are transient in nature but they are 
registered in the subconsciousness and may become noticed in later conscious scenarios.

+ Note that entropy terms behave differently from the usual energy function of Euclidean distance in artificial neural networks. The attractors in 

+ noise provides the insight (Monte Carlo computation). ``free will'' is thus interpretation as internal randomness. 
}
\hide{
\section{Discussions on CER}
There is a subtle difference between consciousness as an evolved mechanism/ability and a state of mind just out of a coma. Human use of the term {\em consciousness} is ambiguous in this regard. For example, a conscious scenario may be  restricted to a narrow sensory channel instead of some integration of multiple channels in a whole scenario, if the brain is impaired in some way or in some non-sober state. That does not negate the existence of integration as a mechanism for consciousness. 

+ Our claim is not that the CER simulation of these reported activities is a fully accurate and detailed depiction of what really happens in our brain. Our hypothesis is that it is advantageous to introduce such a mathematical model in which conceptual ideas can be tested, manipulated and computer-simulated. 

+ we are not suggesting entropy as an mechanism for selection. it is the mathematical description of the mechanism, which is unspecified or unknown. But we know such a selection is possible in neural networks. gradient descent is performed on combined states not the channels, non Turing

+  The experience of the blue square cannot be reduced to experiences of
its components, i.e., an experience of the blueness and an experience of the squareness.---- Scenario vs individual patterns

+ By defining consciousness as a mechanism instead of some numerical measurement, we avoid the trouble of demarcating one part of the brain for consciousness one day and a different part the next day.

+ The mysterious feelings surrounding ``subjective experiences'' are illusional.

\section{Talk Abstracts}
Consciousness as Entropy Reduction, Part 1

Abstract [200 words]

Though familiar to all, `consciousness' lacks a definition. One difficulty is the variety of interpretations across various disciplines with contrasting approaches. Another is the need for a third-person account of subjective experience. Furthermore the quest includes the meaning of consciousness for entities other than humans, including AIs; and the notion has not been clarified by thoughts about related concepts like self awareness, empathy and mental capacity.

In this and the following talk, a model of consciousness is considered which, having a logical basis, lends itself to simulation using a simple mathematical model called Consciousness as Entropy Reduction (CER). The work owes much to GWT, IIT and the earlier less mainstream `feature-map model' in Psychology.

CER considers the contents of consciousness and subconsciousness as `scenarios': vectors of patterns (or features) on various channels (or feature locations). A feature map is one subconscious input scenario amongst many from which the conscious experience is chosen. The result is an internal simulation of the outside world, as in `predictive processing' from Neuroscience. Solving problems internally by simulation is more efficient than experimenting in the actual environment. Conscious experience in such a form is not only efficient but amenable to adaptability, a major evolutionary advantage.

******************************************************************************

Consciousness as Entropy Reduction, Part 2

Abstract [200 words]

In Consciousness as Entropy Reduction (CER) subconscious content is modelled as a probability distribution of features over channels. A mechanism chooses one for consciousness by gradient descent of the entropy of the distribution until a zero-entropy deterministic distribution is reached. CER offers the advantage of simulation.

The brain is seen as an entropy-reducing mechanism that keeps `squeezing' the contents in the subconsciousness to form scenarios of conscious experience. If there is an information pathway that projects conscious experiences back to the subconscious, the mechanism may produce consequent experiences from previous experiences, giving rise to `high-order' thoughts---a characteristic of more evolved forms of consciousness. CER does not distinguish the objective from the subjective. Subjective concepts such as self, self awareness, self consciousness and subjective experiences become internally generated abstract patterns appearing in conscious scenarios.

The gradient descent of entropy reduction in CER resembles that of a typical Hopfield neural network, which changes state by energy reduction until it reaches an energy-minimal stable state. In CER the entropy of a subconscious distribution may be considered `energy' and its stable states the extreme distributions with zero entropy.

We analyse: the experiment of binocular rivalry from human vision; Libet's experiment from Neuroscience; thinking fast and slow; and the formation of higher-order consciousness.

\newpage
\subsection{ExtendedAbs}
+ A model of consciousness is proposed which, having a logical basis, lends 
itself to simulation, as demonstrated on the familiar benchmark of binocular 
rivalry. The model derives from the ideas of Predictive Processing or Coding, 
in which a sentient agent acts according to an internal model of its 
environment, and an associative (Hopfield-like) memory structure with 
feedback, whose detail determines the accuracy of the internal model and 
hence the richness of the agent's choice of action. Adaptation (in the short
term) and evolution (in the much longer) are seen as entropy decreasing, 
and used to account for properties of the model: ...
structure  which assigns an order (as in higher order) to events. 

+ Consciousness is a concept that is familiar to everyone, but a precise and consensual definition of consciousness is still not within reach. Not only because of our limited knowledge about the underlying neural mechanism, but also because we do not really agree on exactly what is  ``the ability of having consciousness''. If we insist on the inclusion of empathy (or ``playfulness'') as required characteristics, then any species such as reptiles without empathy would not be considered to possess consciousness. On the other hand, if we consider any species with the ability of dreaming (or imagining) to have consciousness, then scientific studies have shown that at least some reptiles indeed dream. Then humans simply have a more advanced form of consciousness than that of reptiles.

\subsection{Motivating  Paragraph}
Draft for: `Which papers or preprints is your presentation based on?'

We have published four papers on a Computer Science approach to consciousness of agents, and a fifth (invited) is currently being refereed; see below. 

This talk starts a new direction, on which we would greatly value responses and feedback from our MoC colleagues in their various disciplines. It combines the evolution of consciousness with entropy-decreasing search and algorithmic game theory.

1. A modal approach to consciousness of agents. 
In \textit{Leveraging Applications of Formal Methods, Verification and Validation, Adaptation and Learning}. Editors T.\,Margaria \& B.\,Steffen. Springer Lecture Notes in Computer Science, {\bf 13703}:1--15, 2022.

2. A modal approach to conscious social agents, 
In \textit{International Journal on Software Tools for Technology Transfer, Foundations for Mastering Change}, Rigorous Engineering of Collective Adaptive Systems. Springer Nature, 10 pages, 2023.

3. Consciousness by degree.
In \textit{Theories of Programming and Formal Methods: Essays Dedicated to Jifeng He on the Occasion of His 80th Birthday}. Editors Jonathan Bowen, Qin Li \& Qiwen Xu. Springer Lecture Notes in Computer Science, {\bf 14080}:87--109, 2023.

4. The evolving conscious agent, I.
In \textit{Leveraging Applications of Formal Methods, Verification and Validation: Rigorous Engineering of
Collective Adaptive Systems, Part II}. Editors T.\,Margaria \& B.\,Steffen. Springer Lecture Notes in Computer Science, {\bf 15220}:88--103, 2024.

5. The evolving conscious agent, II.
Submitted to \textit{International Journal on Software Tools for Technology Transfer,
Foundations for Mastering Change}, 
Rigorous Engineering of Collective Adaptive Systems
Springer Nature, 2025.

\section{MISC}
+ similar to regional extinction events and breakthrough events evolution, a squeezed information diversity helps cleanse the obsolete historic designs and set the stage for future evolution.

+ $E$ external input, $X$ is the current input. $X_t=E_t$

+ subconsciousness S tracks E, without new conscious experience,S will eventually be E ($\alpha>0.5$):
\[S_{t+1} = (S_t\;+_\alpha\; X_t)\;+_\epsilon \;\overline{C_t} \]

+ \[C=grediantdescent(S)\]

+ conscious patterns C is removed from S, make sure the less a conscious pattern is in the SUb, the less it is removed. void is also removable from the options to increase sensitivity.

\section{Related work}
\subsection{Eigen, 1971}

Manfred Eigen \cite{eigen} gave an early physical interpretation of 
`survival of the fittest' from the viewpoint of a biophysical 
chemist. He stressed the 
importance of noise in producing random exploration by 
mutation, the dominant view at the time before the later 
extra-genetic advances of evolutionary biology. His approach 
was based on the role of information, being given meaning 
through the functionality it facilitates. The relevance for us is
that led him to use entropy.
\begin{quote}
\textit{
 Fitness is dependent upon constant fluxes of entropy, leading 
 to instability and selective advantages, leading to evolution.} 
\hspace*{\fill}Eigen, \cite{eigen}.\\
\end{quote}
\vspace{-2ex}

In the setting of our recent paper \cite{II} that translates to 
exploration of `fitness space' \textfrak{F}, and hence to the
current approach of CER.

\subsection{Libet, 1983}
In the first half of the 1980s, neuroscientist Benjamin Libet 
performed a series of experiments comparing the timing of 
a participant's mental readiness potential with the timings 
of its decision to act and actual action (see Libet \textit{et al.}, 
\cite{libet}).

The experiment consisted of a participant seated before an
oscilloscope on which a dot rotated quickly with constant 
speed. The participant noted the position of the dot at the
instant `he/she was first aware of the wish or urge' (action 
$B$) to push a button (action $C$). From electrodes 
attached to the participant's brain, Libet discovered that
the instant of the related neurological impulse (action $A$) 
preceded action $B$ by a surprising interval. 
See Fig.\,\ref{fig:libet}. In his words:
\begin{quote}
It is concluded that cerebral initiation of a spontaneous, freely voluntary act can begin unconsciously, that is, before there is any (at least recallable) subjective awareness that a ‘decision’ to act has already been initiated cerebrally. This introduces certain constraints on the potentiality for conscious initiation and control of voluntary acts. 
\hfill Libet \textit{et al.}\ \cite{libet}
\end{quote}

\begin{figure}[ht]
\begin{center}
\begin{picture}(360,40)(0,0)
\thicklines 
\put(0,0){\vector(1,0){280}} 
\put(284, -3){\footnotesize \textit{time\ in\ millisecs}}
\put(30,0){\vector(0,1){40}}
\put(16,-12){\footnotesize \textit{-300}}
\put(26,-23){\footnotesize $A$}
\put(10,42){$\overbrace{\hspace*{1.4cm}}$} 
\put(20,51){\footnotesize $+^\circ\textit{50}$}
\put(150,0){\vector(0,1){40}}
\put(148,-12){\footnotesize \textit{0}}
\put(146,-23){\footnotesize $B$}
\put(130,42){$\overbrace{\hspace*{1.4cm}}$}
\put(140,51){\footnotesize $+^\circ\textit{50}$} 
\put(230,0){\vector(0,1){40}}
\put(222,-12){\footnotesize \textit{200}}
\put(226,-23){\footnotesize $C$}
\put(210,42){$\overbrace{\hspace*{1.4cm}}$} 
\put(220,51){\footnotesize $+^\circ\textit{50}$}
\end{picture}
\end{center}
\vspace{2ex}
\caption{The temporal separation of events in Libet's experiment, 
in milliseconds, with error of $+^\circ 50$. \fbox{One-sided $-50$?} 
In practice the spikes are Gaussian. $B$ is the conscious 
decision to act; $C$ is the consequent action; and $A$ 
is the preceding neurological impulse.}\label{fig:libet}
\end{figure}

Libet's experiments have provoked work in several directions. 
One concerns the timing for different kinds of action.  For 
instance Maoz \textit{et. al.}\ \cite{moaz} have argued that action
$C$ is `purposeless, unreasoned, and without consequences'
or, in their terms, `arbitrary'. They investigated the degree to 
which that timing of the readiness potential extends to other kinds, 
in particular what they refer to as `deliberate decision-making.' 

They confirmed Libet's experiments for arbitrary decisions,
but found:
\begin{quote}
\ldots\ they were strikingly absent for 
deliberate ones. \ldots\ 
%
\ldots\ We directly compared deliberate 
(actual \$1000 donations to NPOs) and arbitrary decisions, and
found readiness potentials before arbitrary decisions, 
but---critically---not before deliberate decisions. This supports the 
interpretation of readiness potentials as byproducts of accumulation 
of random fluctuations in arbitrary but not deliberate decisions and 
points to different neural mechanisms underlying deliberate and 
arbitrary choice.\\
\hspace*{\fill} Moaz \textit{et al}.\ \cite{moaz}.\\
\end{quote}

Suppose we identify `arbitrary decision' with one involving 
an immediate reaction, and `deliberate decision' with one 
requiring reasoning. We might think in Kahneman's terms 
\cite{kahneman}, where the former matches `thinking fast', 
and the latter `thinking slow'. Is it possible that Kahneman's 
dichotomy actually matches those two kinds of decision? 
Further experimental data are necessary. One interesting
example might be provided by the directional decisions 
made by rats in a maze, due to their grid cells \cite{moser}.

\subsection{Baars, 1998 and refinements}
Bernard Baars' Global Workspace Theory (GWT), \cite{gwt}, in 
any of its versions, has been hugely influential as an architecture 
for the promotion of subconscious to conscious. 

\begin{quote}
GWT uses the metaphor of a theatre, with conscious thought being 
like material illuminated on the main stage. Attention acts as a spotlight, 
bringing some of this unconscious activity into conscious awareness 
on the global workspace. \ldots\ 
The stage receives sensory and abstract information, but only events 
in the spotlight shining on the stage are completely conscious.
\hspace*{\fill}Wikipedia
\end{quote}

\subsection{Feature Integration Theory}
+ [CYF: why is FIT a refinement of GWT?] An important refinement of GWT is that of Predictive Processing 
(or Predictive Coding) in which the agent exploits awareness of
its environment (physical, social and mental) to form a model of 
it based on which it 
chooses its next behaviour (provided the external environment 
does not supervene). That approach has found support from 
Neuroscience, where it arose (see for instance Anil Seth 
\cite{seth}), through Philosophy of the Mind, to AI where it is 
exploited (see for instance Yoshua Bengio \cite{bengio}).

Predictive Processing was preceded by the more specialised
Feature Integration Theory (FIT) in.  Vision (reviewed in 
\cite{aggelo}) developed in Psychology in the 1980s, allowing 
an individual to make inferences from visual perceptions. 

\begin{quote}
Perceptual inference refers to the ability to infer sensory stimuli 
from predictions that result from internal neural representations 
built through prior experience. \ldots\
In this framework, 
perception can be seen as a process qualitatively distinct from 
sensation, a process of information evaluation using previously 
acquired and stored representations (memories) that is guided 
by sensory feedback.

The feature integration theory \ldots\
proposes that different attentional mechanisms
are responsible for binding different features into consciously experienced
wholes. The theory has been one of the most influential psychological models 
of human visual attention. \ldots\
in a first step to visual processing, several primary visual features are processed 
and represented with separate \textit{feature maps} that are later integrated into 
a “saliency map” that can be accessed in order to direct attention
to the most conspicuous areas.
\hspace*{\fill}Aggelopoulos, \cite{aggelo}\\
\end{quote}

FIT was in turn preceded, remarkably in the 19th century, by 
Helmut von Helmholtz's suggestion (see for example his 
\textit{Handbuch} \cite{HvonH} whose three parts were devoted 
to the eye as an optical instrument; the sensations of vision; and 
perception) that visual perception is composed of:
\begin{quote}
a hypothesis about what
is being seen based on “inductive inferences” gained from “sensations”. \ldots\ 
This \ldots\ 
regards perception not primarily as a sensory phenomenon but as
perceptual inference relying on internal models built through past experience.
Helmholtz's idea of perceptual inference has been revived by computational
models of perception relying on statistical inference \ldots
\hspace*{\fill}Aggelopoulos, \cite{aggelo}\\
\end{quote}

That approach has been extended by Hoffman to Computational Evolutionary 
Perception (CEP), \cite{hoffman}, which emphasizes \ldots.

It should be mentioned that, by comparison, \textit{feature maps} 
in neural networks 

\begin{quote}
\ldots are fundamental outputs generated by the 
layers within a Convolutional Neural Network (CNN), 
particularly the convolutional layers. They represent learned 
characteristics or patterns detected in the input \ldots

Feature extraction is the overall process of transforming 
raw data into numerical features, and feature maps are a 
specific type of representation generated during this 
process in vision models. 
\hspace*{\fill}Ultralytics Yolo11, `Feature Maps'.
\end{quote}

\subsection{IIT, 2004}

From 2004 when Giulio Tononi first proposed Integrated 
Information Theory (IIT), \cite{iit}, to its latest version, 
\cite{iit2}, IIT has been a controversial approach to 
consciousness. A summary, with arguments for and against, 
is provided on {\tt Wikipedia} \cite{iit3}.

\subsection{Kahneman, 2011}

Daniel Kahneman \cite{kahneman} identifies two human cognitive 
abilities: `thinking quickly' (and intuitively), and `thinking slowly' 
(by reasoning). Nowadays the first is seen as the form of
reasoning performed by LLM, raising the question of whether or
not they can be made capable of the second.

\subsection{Dehaene, 2014}
Insistence on falsifiability leads to consciousness being 
described in terms of some externally observable behaviour. 

Stanislas Dehaene, \cite{dehaene}, ensures that by
expressing a human to be conscious of those phenomena 
of which it is able to report. For other sentient agents that 
needs an extension  with `report' being replaced by some 
form of body language or social interaction. 

The probabilistic nature of our model if interpreted that way
means that falsifiability of its outcomes is statistical in nature.

\subsection{Hopfield, 1982}

In the early 1980's John Hopfield introduced the networks 
whose generalisations now bear his name \cite{hopfield}. 
He built on several decades of attempts to construct
associative memory: a structure  able to recall stored 
previous input similar to the present input. For instance a 
Hopfield network can be trained to return the correct data 
associated with a similar noisy input. 


A \textit{binary Hopfield network with discrete 
time and $N$ nodes} may be seen as a Recursive Neural 
Network (RNN) incorporating discrete time $t : \mathbb{N}$ 
which is a complete symmetric graph whose nodes 
$j : [0,N)$ have a Boolean activity value $\textit{act}_j$:
\begin{eqnarray*}
\textit{act}_j (t{+}1) & := &
\left\{\begin{array}{rll}
1 & \mbox{if}\quad \sum_{k\neq j} w_{j,k} \textit{act}_k (t) + \textit{in}_j\ >\ 0 \\
0 & \mbox{otherwise} \, ,
\end{array}\right.
\end{eqnarray*}
where 
$\textit{in}_j$ is $j$'s current input, and
$w_{j,k} = w_{k,j}$ is the weight on the symmetrical edge between 
$j$ and $k \neq j$.

In other words, the state $V$ of the network at any time consists of  
the active nodes, given by:
\begin{eqnarray*}
&& V : [0,N) \fun \mathbb{B} \\ 
&& V\!(j) := (\textit{act}_j = 1) \, .
\end{eqnarray*}
In the network's learning stage, the weights are learnt by 
\textit{Hebbian association}, defined in terms of $V$:
\begin{eqnarray*}
w_{i,j} & := & (2V\!(i) {-} 1).(2V\!(j) {-} 1) \, ,
\end{eqnarray*}
after which they remain unchanged. In the absence of a global clock
the network is updated node-by-node:
\begin{eqnarray*}
&& {\bf for\ each\ node}\ j: [0,N) : \\
&& \ \ \ {\bf if}\ (\sum_{k\neq j} w_{j,k} \textit{act}_k  + \textit{in}_j)\, > 0 : \\
&& \ \ \ \ \ \ \textit{act}_j := 1 \\
&& \ \ \ {\bf else} : \\
&& \ \ \ \ \ \ \textit{act}_j := 0 \, .
\end{eqnarray*}

The \textit{energy} (or Lyapunov function,
whose theory, \cite{elaydi} Section 4.5, is used in proofs) 
of a Hopfield network is a real-valued function 
\begin{eqnarray*}
E & := &
 \sum_{j\neq k} w_{j, k} \textit{act}_j \textit{act}_k - \sum_{j} \textit{in}_j \textit{act}_j \, .	
\end{eqnarray*}
$E$ is bounded below and as the network is updated 
$E$ weakly decreases and the network converges 
to a local minimum of $E$ where $V$ is the associated
stored memory. For instance (from Wikipedia \cite{wikihop},
Section training) if the training of a Hopfield network with
$n = 5$ includes $(1,0,1,0,1)$ then input $(1,0,0,0,1)$
may result in gradient descent which recalls $(1,0,1,0,1)$.

\subsection{Bengio, 2019}
Yoshua Bengio has proposed features required for 
consciousness by a learning agent, the consciousness 
prior (though devoid of probabilities normally associated 
with the term \textit{prior} since they are simply beyond 
current knowledge). He ensures consciousness is 
falsifiable by taking it to mean `reportable' in Dehaene's 
sense, and aligns it with the second of Kahneman's 
cognitive abilities. Architecturally, he follows GWT and 
Predictive Processing in particular.

Bengio's model, informed by machine learning, is constructed 
from unconscious states, some of which are promoted to 
conscious states, plus memory, expressed as random variables 
subscripted by time, as follows. 
Unconscious state at any time 
\begin{eqnarray*}
u_t & = & F(u_{t-1},a_t)
\end{eqnarray*}
is given by some function $F$ (the encoder) of previous
unconscious state and features $a_t$ of which the agent 
is currently aware. It can be thought of as an RNN.
Conscious state  
\begin{eqnarray*}
c_t & = & C(c_{t-1}, u_t, \textit{noise}_t, m_{t-1})
\end{eqnarray*}
is given by some function $C$ (the consciousness process)
of previous conscious state, current unconscious state, 
randomly generated noise, and previous memory 
\begin{eqnarray*}
m_t & = & M(m_{t-1}, c_t)\, , 
\end{eqnarray*}
which in turn  is updated by function $M$ from previous 
memory and current conscious state.

The functions $F$, $C$ and $M$ structure the theory with 
the understanding that currently their identification, like that 
of the distributions, is simply infeasible. In reality $F$ and 
$C$ incorporate goals, in which case they are viewed as 
searches. The current paper is devoted to one mechanism
for the search $C$: by entropy reduction.

The random variables form a factor graph whose factors 
capture the strength of dependency between the variables. 
Bengio stresses that an efficient implementation demands 
a sparse factor graph and good representation of the $u_t$ 
to enable abstract explanatory factors to be disentangled.


\section{Model of Conscious Experiences (MCE)}
\subsection{Definitions}
\subsection{An Example: Two-channel Pattern Recognition}
+ Consider a system with two channels $a_0=0.6$ and $b_0=0.7$. Then define their joint (independent) distribution $d\Df a\times b=(a_0b_0,\,a_0b_1,\,a_1b_0,\,a_1b_1)=(0.42,0.18,0.28,0.12)$ as the initial distribution of gradient descent. In general a distribution has the form $d=(x,\,y,\,z,\, w)$  where $x+y+z+w=1$ and $H(d)$ is the refusal entropy of the given distribution. 
Note that the entropy of joint distribution is the sum of entropy values: \[H(x+z)+H(x+y)= E(x+z)+E(y+w)+E(x+y)+E(z+w)= E(x)+E\,y+E(z)+E(w)\]

+ $B_0$ {\em is a recognition of} $A_0$ (written $A_0= B_0$) iff $A_0$ leads to $B_0$ and $B_0$ recalls $A_0$. Two combined states are in favour: $A_0B_0$ and $A_1B_1$, indicating some kind of synchronization. For patterns of two channels, recognition works in both ways symmetrically. For three channels, pattern recognition is a synchronization between $A_0B_0$ (or $\overline{A_0B_0}$) and $C_0$ (or $\overline{C_0}$)
where $\alpha$ is a scaling parameter, representing the strength of the memory, and the scaling factor $e$ ensures.

+ The refusal entropy to minimize is as follows:
\[\mathcal{E}(d)\Df H(d\# A_0)+H(d\# B_0)+\alpha F(d\# A_0\Leftrightarrow B_0)\ =\ H(x+z)+H(x+y)+\alpha F(x+w)\]
\[-\frac{\partial \mathcal{E}(d)}{\partial x}\ =\ \log(e)+\log x-\alpha\log(x+w)\]
\[-\frac{\partial \mathcal{E}(d)}{\partial y}\ =\ \log(e)+\log y\]
\[-\frac{\partial \mathcal{E}(d)}{\partial z}\ =\ \log(e)+\log z\]
\[-\frac{\partial \mathcal{E}(d)}{\partial w}\ =\ \log(e)+\log w-\alpha\log(x+w)\]

\subsection{Simulation}
+ optimization seeks to reduces the entropy by adjusting $x$, $y$, $z$ and $w$. Adjusting vector looks like $(-\beta\frac{\partial \mathcal{E}(d)}{\partial x}, -\beta\frac{\partial \mathcal{E}(d)}{\partial y}, -\beta\frac{\partial \mathcal{E}(d)}{\partial z}, -\beta\frac{\partial \mathcal{E}(d)}{\partial w})$ where $\beta$ is a adjustment rate.

+ Consider $\#0.4\times \#0.7$ and $\#0.2\times \#0.7$, and $\alpha=1$.
The former ends up in $(1,0,0,0)$ (high-probability synchronization creating conscious patterns), while the latter ends in $(0,0,1,0)$ (low-probability synchronization fading away).

\section{Model of Conscious Processes (MCP)}
\subsection{Distributions}
+ $X_i$ the set (including {\tt void}) of patterns at channel $i$. Channel distribution $x_i:X_i\rightarrow [0,1]$ at the channel. $X_1\times X_2\times \cdots\times X_n$ the set of scenarios of $n$ channels. 

+ A scenario distribution $d:D\rightarrow [0,1]$. Such a distribution is Cartesian or inter-channel-independent, if it is the cross product of some channel distributions. We use $a,\ b,\ c,\ e\cdots$ to denote Cartesian scenario distributions, but $d$ for general ones. And $a_i$ is the $i$-th channel distribution.

+ Let $\alpha$ be a $n$-dim vector such that $\alpha_i\in[0,1]$. Two Cartesian distributions $a$ and $b$ are linearly combined as 
\[C= a\oplus_\alpha b\]
where $c_i=\alpha_i\,b_i+(1-\alpha_i)\,a_i$. 

+ A distribution is deterministic if it is a point distribution. Determinism implies  Cartesianism.

\subsection{Formation of Consciousness}
+ A Markovian transition matrix is a matrix with column vectors being Cartesian scenario distributions. It is {\bf Markov-Cartesian} if all column distributions are Cartesian.

+ starts from Cartesian $e$ (as external sensory distribution), entropy-reduced to a deterministic $c$ (as conscious experience), which is then transformed by a Markov-Cartesian matrix $A\,c$ (as subconscious mental associations).

+ entropy $\mathcal{E}$ of a scenario distribution $d$,  scenario distribution $\partial\mathcal{E}/\partial d$. We use $\mathcal{E}[d]$ to denote the limit deterministic distribution of the entropy-reducing process.

+ MCP: $e_0$ the initial Cartesian external distribution, $c_0=\mathsf{void}^n$ the initial (void) conscious experience, entropy $\mathcal{E}$ that determines the formation of consciousness. then the linearly combined subconscious world is as follows: 
\[b_{n+1}\;=\;e_n\oplus_\alpha c_n.\]
The parameter $\alpha\in[0,1]$ represents how much reality contributes to the conscious experience. During sleep, $\alpha$ is close to 0. Dream are almost entirely determined by $A$. Note that in sleep vision and touching may have different levels of sensitivity.

+ the conscious experience formed from $b_{n+1}$ by gradient descent:
\[c_{n+1} \;=\; \mathcal{E}[b_{n+1}].\]
An assumption is adopted that only during the formation of conscious experience  from subconscious scenario distribution, non-Cartesian distributions are involved.

\section{MCE Examples}
+ disclaimer: not about what really happens in brain, but about what could have happened to the simulation model in mathematics.

\subsection{Action}
+ when threshold of prob of an action reaches 0.5, the action is triggered

+ that means either an unconscious association between sensory channel and action channel triggers an action or a conscious one does so by 

\subsection{Simple Maze (Greedy Search))}

\subsection{Blind Sight and Unconscious Reactions}
+ obstacle or any item, a probability distribution, 

+ blind sight does not lead to conscious choice

+ instead when the signal exceeds a threshold, it triggers a pattern for preemptive  action. 

+ cocktail part example (attention is competition between conscious and unconscious)

\subsection{Time Discrepancies in Reaction: Finding Scissors}
+ attention works by prioritizing the gradient descent for rules

+ unlike IIT, we assume the brain works on local distribution, not distribution on all combined states

+ consciousness cannot access subconscious content directly. 

+ {\tt scissors} as a denotation appears at an internal channel for individual objects. 

+ realistic patterns to trigger key pattern, diff from, key triggers memorized scene, too much assumption

+ d1 is the  visual scene of the shelf and an internally generated  object to match, c1 is the initial conscious experience that recognizes items at different channels, what its does is to see if any of A0 A1 ... An is the same as the intended object.  non-matching is also a reaction.  this reaction prompts to shift to another strategy in solving the problem: look elsewhere.

+ d2 is a delayed conscious experience derived from d1, but for some reason, d2 results in a recognition (perhaps conscious output satisfies relevant subconscious assumptions cause a decrease). 

+ the recognition invokes the short-term memory to restore scenario,this is the direction from recognition to scenario, which reappears after looking elsewhere

\subsection{CHOOSE: Libet experiment}

\subsection{CHOOSE: Meeting Scenario  and Self Consciousness}
+ the experience of self as an illusion

\section{MCP Examples}
\subsection{New Writing 2025.6.1}
+ Binocular rivalry is a crucial psychological experiment which suggests how the brain works when a conscious experience is formed. 

+ When both left and right eyes are presented with the pictures of a house and a human face separately, the subject neither reports seeing overlapped house and face nor report seeing a jigsawed picture, instead, the subject repeatedly sees each picture in alternation. Subsequent experiments explored the influence of different factors on the frequency and bias of the alternation.

+ The experiment suggests that the conscious experience at any moment is a selected version from multiple possible (subconscious) experiences. The actual mechanism of selection is unclear. A (current) conscious experience may receive less attention during the next conscious experience. 

+ In some sense, the brain reduces the uncertainty of subsciousness and always leads to a certain (less probablistic and less nondeterministic) conscious experience. In other words, the brain reduces entropy when forming a new conscious experience.

+ Our treatment of the phenomenon makes the following simple assumptions [][][] A computer simulation based on the nodel of these assumptions produces a similar pattern of alternating conscious experiences.

\subsection{CHOOSE: Binocular Rivalry}
+ subconscious signals fade away over time if they appear in consciousness. it means the desire to be conscious is satisfied.

+ attention directed by conscious action raises signal across an area of channels

+ One channel $VC=(house, face, void)$ for center of sensory vision,  $d_0$ has slightly higher likelihood of a house,  conscious output is $s_1 = house$, $d_2$ has half of the original likelihood for house, same level of face, we then have $s_2 = face$.

+ imagined visual channels are not sensory visual channels

\subsection{Complex Maze (with Map Building)}
\subsection{Turing Equivalence}
+ the setup of a network that is Turing equivalent

\subsection{Time Discrepancies in Reaction: Finding Scissors}
+ attention works by prioritizing the gradient descent for rules

+ unlike IIT, we assume the brain works on local distribution, not distribution on all combined states

+ consciousness cannot access subconscious content directly. 

+ {\tt scissors} as a denotation appears at an internal channel for individual objects. 

+ realistic patterns to trigger key pattern, diff from, key triggers memorized scene, too much assumption

+ d1 is the  visual scene of the shelf and an internally generated  object to match, c1 is the initial conscious experience that recognizes items at different channels, what its does is to see if any of A0 A1 ... An is the same as the intended object.  non-matching is also a reaction.  this reaction prompts to shift to another strategy in solving the problem: look elsewhere.

+ d2 is a delayed conscious experience derived from d1, but for some reason, d2 results in a recognition (perhaps conscious output satisfies relevant subconscious assumptions cause a decrease). 

+ the recognition invokes the short-term memory to restore scenario,this is the direction from recognition to scenario, which reappears after looking elsewhere

\subsection{CHOOSE: Meeting Scenario  and Self Consciousness}

\subsection{High Order: Instinct vs Belief Code}
+ geometry vision, functional programming, how shame works

\subsection{Game-Theoretic Reasoning}
+  make it one single example of  game theory, as it is widely treated in other works and not typical to philosophers 

The authors have previously considered agents in a social context
by modelling an agent's `empathy', \cite{snat}. They examined:
The Sally-Anne Test, a clinical test for autism \cite{sat}; and 
The Keynesian Beauty Contest and the market, 
\cite{keynes,soros}. An agent using empathy to anticipate the 
behaviour of another, as in The Sally-Anne Test, is demonstrating
anticipation to depth 1. An agent doing so by considering the
other agent to be doing the same thing and anticipating its 
behaviour, is demonstrating anticipation to depth 2; and so on.
In the Theory of Mind that hierarchy is referred to as higher-order
\cite{stanford}. 

In this section we consider a humorous example of higher-order 
reasoning to different depths from 
\textit{The Romance of The Three Kingdoms}, (\textit{San Guo Yan Yi}),
\cite{britannica,sanguo}, one of China's classical epic novels.

At one point in the novel, Cavalry Commander Cao Cao is fleeing 
after being routed at sea. He finds himself beside thick woods and, 
to the amazement of his generals, begins laughing. He explains
that his opponent Commander Zhou Yu must be witless for not 
laying an ambush there. Next moment they are ambushed by Zhou 
Yu's men. In our terms, while Cao Cao simply acted, Zhou Yu 
anticipated his behaviour by thinking to depth (at least) one.

After escaping, Cao Cao comes to a fork in the road. One branch
is the main road to Yiling and the other a mountain pass to north 
Yiling. He takes the more difficult mountain pass on the grounds
that Zhou Yu would expect him to take the easier main road and
at some point again begins to laugh at having outwitted his 
opponent. Next moment Zhou Yu's ambush strikes. Again, Zhou
Yu has thought one level deeper, to depth (at least) two.

Once more Cao Cao escapes and again is confronted by a fork. 
He has to choose between a longer smooth main road and a shorter
backroad through the treacherous terrain of Huarong Pass. He
regards smoke on the backroad as a trick by Zhou Yu to entice him 
to choose the main road, so chooses the backroad. Once again in
spite of losing many of his troops to the terrible terrain he chortles 
at the stupidity of his opponents for not ambushing his meagre
company of 300. Just then he is ambushed by 500, set by Zhou
Yu yet again thinking one level deeper.

Those cameos from \textit{San Guo} provide entertaining examples
of human agents making decisions at various levels. Confronted 
with a fork in the road Cao Cao must make a choice. Before doing 
so his entropy in \textit{decision space} is maximal, but having 
chosen it reduces to 0. 

The agent-independent means used in our previous work to 
describe that decision is by an agent's \textit{policy} which chooses 
its next action on the basis of an internal model of its environment 
(including other agents). The policy, in tandem with the model, does 
so to a certain depth of anticipating the opponent. In the first cameo, 
Cao Cao's uses depth zero whilst Zhou Yu's uses depth at least one. 

Further elaboration requires further detail of decision space. 

Given the type $\mathbb{T}$ of the system, 
decision space has type $\mathbb{T} \times \mathbb{R}^{\geq 0}$ 
where the second component measures entropy. For instance if
Cao Cao is equally likely to take either road, $\# \mathbb{T} = 2$
then the curvature may be depicted, with $0 = main\, road$ and 
$1 = mountain\, road$:
\begin{center}
\includegraphics[scale = .25]{graph.eps}
\end{center}
\hfill {\footnotesize {web image}}\\
Before deciding Cao Cao is represented by the point with 
maximum entropy and after by either point with zero entropy. 

Thus decision space is endowed with space-time-like curvature 
in which the policy's decision results in gradient descent from 
maximum entropy to zero, with a deterministic decision.

\fbox{Metric:}\quad $l^1$ or $l^2$ or K-L?


\section{Discussions}

+ compare to GWT and Blums' work, our model also `chooses' some `global work-space content' (or a global scenario), but it does not choose it through a tree-structure of comparisons, instead, it chooses over a Hopfield-Network-like entropical gradient-descent process.

\hide{\section{Patterns}
+ The refusal entropy to minimize is as follows:
\[H(d)\Df (H_\top(d) +  H_{A_0,B_0)/\ln 2}(d)\ =\  (x+z)\log\frac{1}{x+y}-x\log x-y\log y-z\log z-w\log w\]
\[\frac{\partial H}{\partial x}\ =\ \frac{x+z}{x+y}+\log(x+y)+ \log w-\log x\]
\[\frac{\partial H}{\partial y}\ =\ \frac{x+z}{x+y}+ \log w-\log y\]
\[\frac{\partial H}{\partial z}\ =\ \log\frac{1}{x+y}+\log w-\log z\]
+ optimization seeks to reduces the entropy by adjusting $x$, $y$ and $z$. $\frac{\partial H(a\times b)}{\partial x}\approx -0.861$, $\frac{\partial H(a\times b)}{\partial y}\approx 1.10$ and $\frac{\partial H(a\times b)}{\partial z}\approx -0.485$. That means to increase $x$, decrease $y$ and increase $z$.

+ In general a distribution has the form $d=(x,\,y,\,z,\, w)$  where $w=1-x-y-z$. Define entropy of the predicate $A_0\Rightarrow B_0$ that corresponds to the states $A_0 B_0,\,A_1 B_0,\, A_1 B_1$. Predicative probability $d[P]=\sum_{i\in P} c_i$:
\[d[A_0]\ =\ x.\]
\[d[A_0\Rightarrow B_0]\ =\ 1-y.\]
\[H_{A_0,B_0}(d)\Df b_0\times KL(\,\# 1,\,\#\, d[A_0\Rightarrow B_0]\,)=(x+z)\log\frac{1}{1-y}\]
where $H_{A_0, B_0}$ is in fact the KL-distance of the distribution $d$ over the predicate  $A_0\Rightarrow B_0$ (and its negation) from the extreme distribution of predicate. There are two ways to reduce this entropy to 0: either by satisfying the predicate or by reducing the credibility  $b_0$ of the rule.

+ if it is KL, the trouble would be when they disagree, entropy becomes infinity. if only positive half, trouble is prob tends to become 0 for all possible pattern recognition. 

+ we need something that can disable a successful rule, but when possible, will recognize the scenario. 

+ how about a0 log- b0 - a0 log- a0? (without a1 part)

+ The refusal entropy to minimize is as follows:
\[H(d)\Df H_\top(d) +  H_{A_0|B_0}(d)\ =\ - (x+z)\log(x+y)-x\log x-y\log y-z\log z-w\log w\]
\[\frac{\partial H}{\partial x}=\log w-\frac{x+z}{x+y}-\log(x+y)-\log x\]
\[\frac{\partial H}{\partial y}=\log w-\frac{x+z}{x+y}-\log y\]
\[\frac{\partial H}{\partial z}=\log w-\log(x+y)-\log z\]
+ optimization seeks to reduces the entropy by adjusting $x$, $y$ and $z$. $\frac{\partial H(a\times b)}{\partial x}=\log(0.12)-\log(0.42)\approx -1.908$, $\frac{\partial H(a\times b)}{\partial y}\approx -1.572$ and $\frac{\partial H(a\times b)}{\partial z}\approx -0.33$. That means to increase $x$, decrease $y$ and increase $z$.

+ consider $N=2$ and $a(1)=0.4$ and $a(2)=0.6$. Then consider $b$ such that $b(1)=1$ and $b(0)=0$, then:
\[Hr(a,b)=\infty\]

\[\msf{P}(\mi{house})\]
a pattern recognized as  a house 
\[\mathsf{L}(\mi{front})\]
front layer of the vision field (default)
\[\msf{G}(\mi{sub})\]
the pattern at the stage of subconsciousness (or consciousness)
\[\msf{O}(\mi{reality})\]
origin of the pattern from reality (or generated action, imaginative intermediate patterns), the brain, when it is in the right state, distinguishes whether a pattern is realistic or imagined
\[\msf{A}(\mi{high})\]
attention high (or low) for some specific kind of patterns

\newpage
\section{Modal Logics}
Modality of subconscious imagination 
\[\Diamond\;P\]
Modality of (partial) ordering of importance, being normal, transitive and idempotent, patterns 
\[\blacklozenge\;P\]
Modality of conscious choice
\[\varDelta\,P\]
Rules of conscious choice
\[\frac{\blacklozenge\;P\rightarrow \blacklozenge\;S}{\varDelta\,P\rightarrow S}\]
\[\blacklozenge\;P\rightarrow \blacklozenge\varDelta\,P\]
Provable laws
\[\varDelta\,P\rightarrow \varDelta\varDelta\,P\]
\[\varDelta\,P\rightarrow P\]

\section{MISC: desire/relation/pain/de-relation}

+ all relationships have some asymmetry

+ distinguish accepted and rejected targets

+ faith desire, rejecting incompatible faith

+ grouping desire, leading and following, leader competition 

+ tribal relation

+ fostering and fostered

+ sexual desire

+ individual safety (avoidance of death and harm threat), reflecting a relation between central nervous system and the rest of the body of an individual  

+ reality, harmonious relation (clear recognition of what is imagination and not reality) between conscious formation and sense of reality (if reality is the source of unbearable pain, a detachment between consciousness and reality sense is  initiated, reality becomes the target of rejection, denial of reality

\section{abstract}

+ functional

+ with feedbacks

+ with options and strength likelyhood, and competition

+ attention as actions

+ local options and combined evaluation and choice

+ with memory of scenarios and auto filling 

+ scenario evaluation and choice

+ feedback of the chosen scenario

+ every variable has an internal representation var?

+ evaluation network

+ $x$ realityy signal from the outside, $\overline{x}$ signal originated from the inside, we may need $x_{lt}$ to indicate the signal at location $l$ (within the entire realityy field) at time $t$.

+ the presence of overlined variables means consciousness in the modern sense

\section{Behavioral Model }
+ assumption: direct behavioral reaction is quick, compared to the formation of consciousness, thus the reaction is regarded to occur ``at the same time'' with subconscious inputs

+ actionable channels $a[t]=a_1[t],a_2[t],\cdots$, subconscious channels $b[t]=b_1[t],b_2[t],\cdots$

+ a scenario is a singleton-set deterministic predicate

+ $S(b)$  a scenario of input channels, but arbitrary for other channels

+ two possible scenarios $S_1\vee S_2$
\[B_t\Df \bigwedge_{i=1}^n\  a_i[t] = f_i(b_{i_1}[t],\cdots, b_{i_m}[t])\]
\[B(a,b)\Df \bigwedge_{t=0}^\infty B_t\]
+ pure behavioral reaction to some scenario:
\[S(b)\wedge B(a,b)\]

+ vacant channel (no object or an object difficult to notice or act upon) $b_i=\uparrow$

\section{Conscious Model}
+ pre-conscious channels $d[t]=c_1[t],c_2[t],\cdots$, conscious channels $d'[t]=d'_1[t],d'_2[t],\cdots$

+ a conscious reaction
\[C\Df C(a,b,d,d')\]
+ conscious model
\[C_1;\; C_2\Df \]
+ wrap up C1's $d'$ and C2's $d$, and use predicate transformer to choose, not yet worked out

\section{Binocular Rivalry}
\[\msf{P}(\mi{house})\wedge\msf{L}(\mi{front})\wedge\msf{O}(\mi{reality})\wedge\msf{G}(\mi{sub})\ @\;t^\geq\]
\[\msf{P}(\mi{face})\wedge\msf{L}(\mi{front})\wedge\msf{O}(\mi{reality})\wedge\msf{G}(\mi{sub})\ @\;t^\geq\]
\[\msf{P}(\mi{house})\wedge\msf{A}(\mi{high})\wedge\msf{O}(\mi{action})\wedge\msf{G}(\mi{sub})\ @\;t\]
\[\msf{P}(\mi{face})\wedge\msf{A}(\mi{low})\wedge\msf{O}(\mi{action})\wedge\msf{G}(\mi{sub})\ @\;t\]
\[\msf{P}(\mi{house})\wedge\msf{L}(\mi{front})\wedge\msf{O}(\mi{reality})\wedge\msf{G}(\mi{csc})\ @\,t{+}1\]
+ conscious choice is simply for the  pattern with higher attention
\[\msf{P}(\mi{house})\wedge\msf{A}(\mi{low})\wedge\msf{O}(\mi{action})\wedge\msf{G}(\mi{sub})\ @\,t{+}1\]
\[\msf{P}(\mi{face})\wedge\msf{A}(\mi{high})\wedge\msf{O}(\mi{action})\wedge\msf{G}(\mi{sub})\ @\,t{+}1\]
\[\msf{P}(\mi{face})\wedge\msf{L}(\mi{front})\wedge\msf{O}(\mi{reality})\wedge\msf{G}(\mi{csc})\ @\,t{+}2\]
\[\msf{P}(\mi{house})\wedge\msf{A}(\mi{high})\wedge\msf{O}(\mi{action})\wedge\msf{G}(\mi{sub})\ @\;t\]
\[\msf{P}(\mi{face})\wedge\msf{A}(\mi{low})\wedge\msf{O}(\mi{action})\wedge\msf{G}(\mi{sub})\ @\;t\]
\[\cdots\cdots\]

\subsection{Maze Playing}
\subsubsection{A Simple Maze}
+ no conflicting perceptions assumed if location is unspecified
\[\msf{P}(\mi{maze})\wedge\msf{O}(\mi{reality})\wedge\msf{G}(\mi{sub})\ @\;t^\geq\]
+ when attention is not specified, it is arbitrary and not a concern.
\[\msf{P}(\mi{dest})\wedge\msf{O}(\mi{reality})\wedge\msf{G}(\mi{sub})\ @\;t^\geq\]
\[\msf{P}(\mi{source})\wedge\msf{O}(\mi{reality})\wedge\msf{G}(\mi{sub})\ @\;t^\geq\]
+ when location is not specified, it is  somewhere in the front without interfering with other patterns
\[\msf{P}(\mi{Paths(source)})\wedge\msf{O}(\mi{imagin})\wedge\msf{G}(\mi{sub})\ @\;t\]
+ patterns  of all possible path segments from source in subconsciousness
\[\msf{P}(\mi{maze})\wedge\msf{O}(\mi{reality})\wedge\msf{G}(\mi{csc})\ @(t+1)^\geq\]
\[\msf{P}(\mi{dest})\wedge\msf{O}(\mi{reality})\wedge\msf{G}(\mi{csc})\ @(t+1)^\geq\]
\[\msf{P}(\mi{source})\wedge\msf{O}(\mi{reality})\wedge\msf{G}(\mi{csc})\ @(t+1)^\geq\]
choice criteria for the route closest to correct route.
\[\msf{P}(path(source,dest))\wedge\msf{O}(\mi{imagin})\wedge\msf{G}(\mi{csc})\ @\;t+1\]
+ solution emerges immediately in one step: $path$ is a complete path from {\it source} to {\it dest}.

\subsubsection{A Complex Maze}
\[\msf{P}(\mi{maze})\wedge\msf{L}(\mi{back})\wedge\msf{O}(\mi{reality})\wedge\msf{G}(\mi{sub})\ @\;t^\geq\]
\[\msf{P}(\mi{dest})\wedge\msf{L}(\mi{front})\wedge\msf{O}(\mi{reality})\wedge\msf{G}(\mi{sub})\ @\;t^\geq\]
\[\msf{P}(\mi{source})\wedge\msf{L}(\mi{front})\wedge\msf{O}(\mi{reality})\wedge\msf{G}(\mi{sub})\ @\;t^\geq\]
\[\msf{P}(\mi{Paths(source)})\wedge\msf{L}(\mi{front})\wedge\msf{O}(\mi{imagin})\wedge\msf{G}(\mi{sub})\ @\;t\]
\[\msf{P}(\mi{maze})\wedge\msf{L}(\mi{back})\wedge\msf{O}(\mi{reality})\wedge\msf{G}(\mi{csc})\ @(t+1)^\geq\]
\[\msf{P}(\mi{dest})\wedge\msf{L}(\mi{front})\wedge\msf{O}(\mi{reality})\wedge\msf{G}(\mi{csc})\ @(t+1)^\geq\]
\[\msf{P}(\mi{source})\wedge\msf{L}(\mi{front})\wedge\msf{O}(\mi{reality})\wedge\msf{G}(\mi{csc})\ @(t+1)^\geq\]
\[\msf{P}(path(source,x_1))\wedge\msf{L}(\mi{front})\wedge\msf{O}(\mi{imagin})\wedge\msf{G}(\mi{csc})\ @\;t+1\]
\[\msf{P}(\mi{Paths(x_1)})\wedge\msf{L}(\mi{front})\wedge\msf{O}(\mi{imagin})\wedge\msf{G}(\mi{sub})\ @\;t+1\]
+ need location here
+ paths from $x_1$
\[\msf{P}(path(x_1,x_2))\wedge\msf{L}(\mi{front})\wedge\msf{O}(\mi{imagin})\wedge\msf{G}(\mi{csc})\ @\;t+2\]
\[\cdots\cdots\]
\[\msf{P}(\mi{Paths(x_k)})\wedge\msf{L}(\mi{front})\wedge\msf{O}(\mi{imagin})\wedge\msf{G}(\mi{sub})\ @\;t+k\]
\[\msf{P}(path(x_k,dest))\wedge\msf{L}(\mi{front})\wedge\msf{O}(\mi{imagin})\wedge\msf{G}(\mi{csc})\ @\;t+k+1\]

\subsection{Blind Sight}
+ no conflicting perceptions assumed if location is unspecified
\[\msf{P}(\mi{walk})\wedge\msf{O}(\mi{action})\wedge\msf{G}(\mi{sub})\ @\;t^\geq\]
\[\msf{P}(\mi{walk})\wedge\msf{O}(\mi{action})\wedge\msf{G}(\mi{csc})\ @(t+1)^\geq\]
\[\msf{P}(\mi{obstacle})\wedge\msf{O}(\mi{reality})\wedge\msf{G}(\mi{sub})\ @\;t+k\]
\[\msf{P}(\mi{evade})\wedge\msf{O}(\mi{action})\wedge\msf{G}(\mi{sub})\ @\;t+k\]
\[\msf{P}(\mi{evade})\wedge\msf{O}(\mi{action})\wedge\msf{G}(\mi{csc})\ @\;t+k+1\]

\subsection{Time Discrepancies in Reaction}
+ mail package (real), need scissors (imagined), on-shelf, or elsewhere, possible plans, 2-step plan
\[\msf{P}(\mi{delivery\_received})\wedge\msf{O}(\mi{reality})\wedge\msf{G}(\mi{sub})\ @\;t\]
\[\msf{P}(\mi{need\_scissors\_to\_open})\wedge\msf{O}(\mi{imagin})\wedge\msf{G}(\mi{sub})\ @\;t\]
\[\msf{P}(\mi{on\_shelf})\wedge\msf{O}(\mi{imagin})\wedge\msf{G}(\mi{sub})\ @\;t\]
\[\msf{P}(\mi{off\_shelf})\wedge\msf{O}(\mi{imagin})\wedge\msf{G}(\mi{sub})\ @\;t\]
\[\msf{P}(\mi{Search\_Plans})\wedge\msf{L}(\mi{action\_plan})\wedge\msf{O}(\mi{imagin})\wedge\msf{G}(\mi{sub})\ @\;t\]
\[\msf{P}(\mi{first\_on\_then\_off})\wedge\msf{L}(\mi{action\_plan})\wedge\msf{O}(\mi{imagin})\wedge\msf{G}(\mi{csc})\ @\;(t+1)^\geq\]
\[\msf{P}(\mi{check\_onshelf})\wedge\msf{O}(\mi{action})\wedge\msf{G}(\mi{sub})\ @\;t+1\]
\[\msf{P}(\mi{scene\_onshelf})\wedge\msf{O}(\mi{reality})\wedge\msf{G}(\mi{csc})\ @\;t+2\]
\[\msf{P}(\mi{check\_onshelf})\wedge\msf{O}(\mi{action})\wedge\msf{G}(\mi{csc})\ @\;t+1\]
\[\msf{P}(\mi{scissors\_unnoticed\_onshelf})\wedge\msf{O}(\mi{reality})\wedge\msf{G}(\mi{sub})\ @\;t+2\]
\[\msf{P}(\mi{check\_offshelf})\wedge\msf{O}(\mi{action})\wedge\msf{G}(\mi{sub})\ @\;t+3\]
\[\msf{P}(\mi{check\_offshelf})\wedge\msf{O}(\mi{action})\wedge\msf{G}(\mi{csc})\ @\;t+4\]
\[\msf{P}(\mi{scene\_offshelf})\wedge\msf{O}(\mi{reality})\wedge\msf{G}(\mi{sub})\ @\;t+4\]
\[\msf{P}(\mi{scissors\_noticed\_onshelf})\wedge\msf{O}(\mi{reality})\wedge\msf{G}(\mi{csc})\ @\;t+4\]
+ time here should be +k +k+l

\subsection{How to include Figures}
\begin{figure}
\centering
\includegraphics[width=0.25\linewidth]{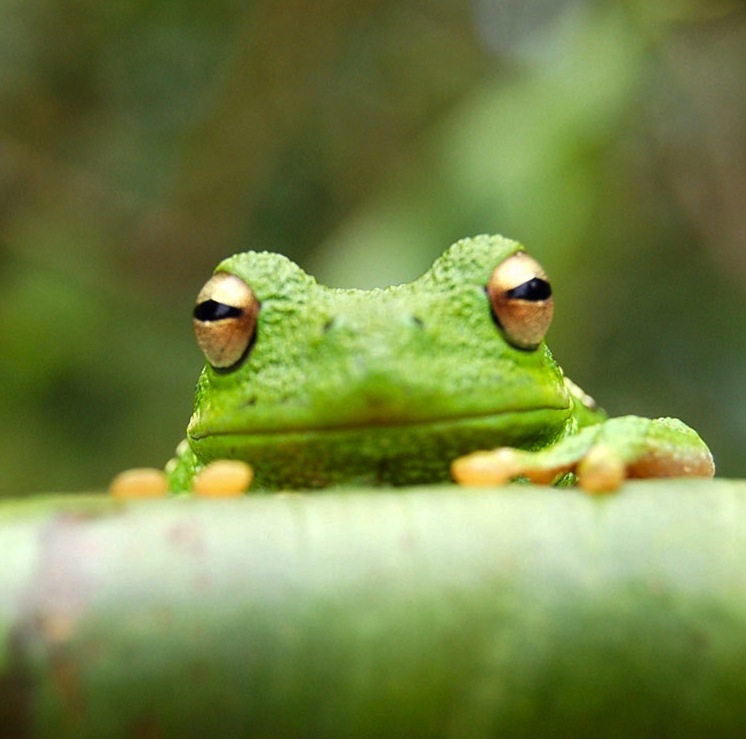}
\caption{\label{fig:frog}This frog was uploaded via the file-tree menu.}
\end{figure}

\subsection{How to add Tables}

\begin{table}
\centering
\begin{tabular}{l|s}
Item & Quantity \\\hline
Widgets & 42 \\
Gadgets & 13
\end{tabular}
\caption{\label{tab:widgets}An example table.}
\end{table}

}
}


\end{document}